# Satellite lines in the photoionization of ions: The Be isoelectronic sequence

W.-C. Chu[1], H.-L. Zhou[1], A. Hibbert[2], and S. T. Manson[1]

[1]*Department of Physics and Astronomy, Georgia State University, Atlanta, Georgia 30303, USA*

[2]*School of Mathematics and Physics, Queen's University of Belfast, Belfast BT7 1NN, United Kingdom*

(Dated: March 17, 2010)

## Abstract

Partial photoionization cross sections from the $^1S$ ground and the $^3P$ metastable states of the Be isoelectronic series to different final states including a number of satellite (ionization plus excitation) channels are studied using the $R$-matrix method. As expected, the ratio of each satellite partial cross section to the dominant partial cross section is found to be a monotonic decreasing function of $Z$ for both the ground and the metastable states. The complicated patterns of these ratios as functions of photon energy are analyzed.

## I. INTRODUCTION

As a basic physical process in nature, the photoionization of atoms and ions has been studied intensively over the years. Since electron-electron correlation plays an important role in the photoionization process in many-body systems, it is of importance to study correlation to fully understand photoionization cross sections. Ionization plus excitation processes, which are usually referred as "satellite lines", are of particular interest since they exist only due to correlation; they are forbidden from a single-particle viewpoint. Thus, investigations of photoionization plus excitation processes constitute an ideal "laboratory" to study correlation in the photoionization process. To date, a number of theoretical and experimental studies of satellite lines have been reported for neutral atoms but very few (especially experimental) for positive ions. It is known, however, that in the limit $Z \to \infty$, satellite lines vanish because the $e^2/r_{ij}$ terms in the Hamiltonian that are the source of correlation become negligible compared with the nuclear Coulomb potential. Thus, while satellite lines for the neutral atom and for infinite $Z$ are known, their evolution with $Z$ has not yet been investigated.

In this paper, a theoretical study of a number of the satellite lines in the Be isoelectronic sequence resulting from the photoionization of the $^1S^e$ ground and the $^3P^o$ metastable states is presented. The 4-electron Be-like systems are relatively simple systems and can be handled quite well theoretically; excellent agreement with measured total cross sections has been found [1]. In addition, these ions are among the abundant species in astrophysical plasmas. This study includes 14 ions, Be, $B^+$, $C^{+2}$, $N^{+3}$, $O^{+4}$, $Ne^{+6}$, $Mg^{+8}$, $Si^{+10}$, $S^{+12}$, $Ar^{+14}$, $Ca^{+16}$, $Ti^{+18}$, $Cr^{+20}$, and $Fe^{+22}$, the most astrophysically abundant of the sequence. To simplify matters, we ignore the region of autoionizing resonances and focus on the photon energy range above the double ionization threshold. In this report we calculate the partial cross section covering the photon energy range up to $8.0 \times E_{range}$, where $E_{range}$ is the energy range from the ionization threshold ($2s$) to the $4f$ threshold for each ion. Each satellite line is compared with the dominant partial cross section and these ratios are examined. How the ratios behave as functions of photon energy as well as how these ratios change as functions of $Z$ are studied. The nonrelativistic $R$-matrix method [2], introduced in the next section, is employed to calculate the partial cross sections. For the photoionization cross sections of these systems, the nonrelativistic and Breit-Pauli

*R*-matrix methods differ mainly in the splitting of resonances and in threshold energies, but the agreement between the nonresonant cross sections predicted by the two methods is quite good [1]. Thus it is adequate to use a nonrelativistic treatment of the photoionization process.

## II. THEORETICAL METHODOLOGY

The photoionization processes for ground and metastable states are given nonrelativistically by

$$1s^2 2s^2 (^1S^e) + h\nu \rightarrow [1s^2 nl + e^-(\varepsilon l')](^1P^o) \tag{1}$$

and

$$1s^2 2s 2p(^3P^o) + h\nu \rightarrow [1s^2 nl + e^-(\varepsilon l')](^3S^e, ^3P^e, ^3D^e) \tag{2}$$

respectively, where the terms are expressed in the *LS* framework. In the *R*-matrix calculation, the so-called target (*N*-electron) discrete wave functions are constructed, and the photoelectron is added onto the target wave functions to construct the initial and final state wave functions of the (*N*+1)-particle system. To build the target state wave functions, we used the 1*s* and 2*s* orbital functions from the Hartree-Fock calculations of the Li-like ground states as given by Weiss [3] and by Clementi and Roetti [4]. With this 1*s* function held in common to all states, we constructed 9 configurations: $1s^2 2s$, $1s^2 2p$, $1s^2 3s$, $1s^2 3p$, $1s^2 3d$, $1s^2 4s$, $1s^2 4p$, $1s^2 4d$, and $1s^2 4f$. The outer *nl* orbital function was optimized, using the CIV3 program [5], on the energy of the corresponding $^2L$ state. The 9 configurations are taken as the basis for the target wave functions, which are calculated by diagonalizing the *N*-particle Hamiltonian.

The core idea of *R*-matrix method is to partition space into internal and external regions, divided by a spherical shell ($r = a$) centered at the nucleus, calculate the wave functions in the two regions with the relevant boundary conditions, by means of the R-matrix at $r = a$. The radius of the sphere *a* is determined by the condition that all the discrete wave functions are small enough at $r = a$ so that they can be neglected in the external region, which means only the continuum extends to the external region. In the internal region, all the *N*+1 electrons are indistinguishable and all exchange terms are taken into account. Thus, the total wave function is given by

$$\Psi_k^{LS\pi} = A\sum_{ij} c_{ijk} \overline{\Phi}_i(S_i L_i; x_1, x_2, \ldots, x_N, \hat{x}_{N+1})\frac{F_j(r_{N+1})}{r_{N+1}} + \sum_j d_{jk}\chi_j^{SL\pi} \qquad (3)$$

where $\overline{\Phi}_i$ are the target wave functions coupled to the angular and spin parts of the photoelectron wave function, the $F_j(r_{N+1})$ are the continuum radial functions of the photoelectron, and $A$ is the antisymmetry operator. Note that the coupling between the target and the photoelectron must match the total angular momentum and the total spin specified by $LS\pi$ in each calculation. The second term in Eq. (3) is a linear combination of discrete total wave functions to ensure the completeness of the expansion. In the external region, the whole system is viewed as a two body system where the exchange terms between the photoelectron and any other electron are small enough to be omitted. The total continuum wave function in the outer region is then given by

$$\Psi_k^{SL\pi} = \sum_{ij} c_{ijk} \overline{\Phi}_i(S_i L_i; x_1, x_2, \ldots, x_N, \hat{x}_{N+1})\frac{F_j(r_{N+1})}{r_{N+1}}. \qquad (4)$$

Note that there is no antisymmetry operator $A$ in Eq. (4). With the forms of Eq. (3) and Eq. (4), by solving the Schrödinger equation with the boundary conditions at $r = 0$, at $r \to \infty$, and the continuity at $r = a$, the wave functions in both regions are obtained using the $R$-matrix as the connection between the regions. Thus, we obtain the initial state wave function with the corresponding initial state energy and the final state wave function with any given total energy over all space. The transition matrix elements are then calculated within the framework of the electric dipole approximation to give the photoionization cross sections.

In the cross section calculation, we consider 400 energy points evenly distributed in the energy range from the ionization threshold to $8E_{\text{range}}$. This energy mesh is good enough to describe the cross sections in this range since we focus on the energy region above the resonances.

## III. RESULTS AND DISCUSSION

For both the ground state and the metastable state, above the resonance region, all the partial cross sections from the initial state to the $1s^2 nl$ final states (designated $\sigma_{nl}$) are smooth functions of photon energy $E_{\text{ph}}$, and each $\sigma_{nl}$ as a function of $E_{\text{ph}}$ also evolves smoothly with Z. Since the energy scale increases with Z, to analyze the evolution along Z in detail, these ratios are plotted against the energy scaled by the factor $E_{range}$, the energy between the 2s and 4f thresholds ($E_{ph}/E_{range}$) of each ion, so that the ratios for different ions are easily compared.

For the photoionization of the $^1S^e$ ground state, $\sigma_{2s}$ is the largest cross section since it results from a single-electron transition. Thus, the ratio $\sigma_{nl}/\sigma_{2s}$ as a function of scaled energy is studied. Figs. 1-3 show our results for the ratios $\sigma_{nl}/\sigma_{2s}$ for neutral Be, $Si^{+10}$, and $Fe^{+22}$, respectively. We first examine the overall strength of each $\sigma_{nl}$ over the whole energy range. As seen in the Fig. 1 for Be, the largest ratio, $\sigma_{2p}/\sigma_{2s}$, varies between 0.1 and 1, while the smallest, $\sigma_{4f}/\sigma_{2s}$, lies between 0.01 and 0.1. For $Fe^{+22}$, which is the highest-Z ion in this study, $\sigma_{2p}/\sigma_{2s}$ varies between 0.01 and 0.1, and $\sigma_{4f}/\sigma_{2s}$ is quite small, of the order of $10^{-5}$. All ratios decrease with Z, which means $\sigma_{2s}$ is more and more dominant at high Z's. This is consistent with what must happen as $Z \to \infty$, and the system becomes hydrogenic; in that limit, only $\sigma_{2s}$ is non-zero and all other $\sigma_{nl}$, which result from two-electron transitions, vanish. In addition, it is seen that at low Z, all the partial cross section strengths clustered together, while at high Z they are more separated.

Looking at the ratios as a function of energy, it is evident from Figs. 1-3 that the ns ratios, $\sigma_{3s}/\sigma_{2s}$ and $\sigma_{4s}/\sigma_{2s}$, are more or less energy-independent (horizontal) at the higher energies, indicating that all of the ns cross sections, including the 2s, have the same energy dependence at the higher energies. We emphasize that transitions to a final ns state have the same energy dependence, at the higher energies, irrespective of whether the transition is a one-electron or a two-electron transition. For all of the transitions to nl states with $l \neq 0$, the ratios fall off with energy, at the higher energies. This shows that ns cross section will be more and more dominant with increasing energy.

To take a closer look at how these overall strengths change with Z, Fig. 4 shows $\sigma_{nl}/\sigma_{2s}$ at $E_{ph} = 8E_{range}$ vs. $1/Z$. The partial cross sections are examined at an energy much higher than all the thresholds, insuring that only the background cross sections, and no resonance structure, are taken into account. As expected, the high-Z ratios tend toward zero because of the hydrogenic nature of high-Z ions. Note that on this $1/Z$ plot, the ratios for higher Z ions that have not been explicitly calculated in this paper are easily obtained by interpolation.

The ratios $\sigma_{nl}/\sigma_{2p}$ for the photoionization of the metastable state for neutral Be, $Si^{+14}$, and $Fe^{+22}$ are shown in Fig. 5-7. For the metastable state, the relative strengths are displayed with respect to $\sigma_{2p}$ since $\sigma_{2p}$ is the largest partial cross section at high energies. However, different from the ground state case, the single-electron ionization process in the metastable state gives rise to both $\sigma_{2s}$ and $\sigma_{2p}$ partial cross sections; thus, as seen in Fig. 5-7, $\sigma_{2s}/\sigma_{2p}$ does not decrease much with Z. On the contrary, the whole $\sigma_{2s}/\sigma_{2p}$ curve stays almost the same from Si$^{+14}$ to Fe$^{+22}$, which indicates the ratio between $\sigma_{2s}$ and $\sigma_{2p}$ is almost constant with Z (at any specific scaled energy) at high Z's, although at really high energy, the 2s cross section will dominate [6]. Other than these two single-electron-excitation partial cross sections, all other cross sections decrease monotonically with Z, just as in the case of the ground state. For the metastable state, the strengths of all $\sigma_{nl}$ are well aligned in the order of *n* for the same *l*. For example, $\sigma_{2s} > \sigma_{3s} > \sigma_{4s}$ for *n*s states. This regularity is mostly true for ground state partial cross sections, except $\sigma_{4p}$ climbs across $\sigma_{3p}$ and is higher than $\sigma_{3p}$ at high energies, as seen in Fig. 2 and 3.

Looking at the cross section ratios as a function of energy, it is noted that all $\sigma_{nl}/\sigma_{2p}$ decrease with energy except $\sigma_{3p}/\sigma_{2p}$ which becomes constant (horizontal) at the higher energies for all three metastable cases. In other words, at the higher energies considered here, the 2p and 3p cross sections have the same energy dependence, but the 4p cross section does not. This differs from the ground state results where all *n*s cross sections exhibited the same energy dependence at the higher energies. The reason for this difference is interchannel coupling; the 4p and 4d cross sections are about the same order of magnitude and their thresholds are quite close in energy. Since the strength of interchannel coupling falls off with increasing energy [7], it is expected that $\sigma_{4p}/\sigma_{2p}$ will become constant at much higher energies. We note further that for metastable state photoionization, the three *n*s cross sections are parallel at the higher energies, i.e., $\sigma_{3s}/\sigma_{2s}$ and $\sigma_{4s}/\sigma_{2s}$ are constant, as a function of energy, at the higher energies; evidently interchannel coupling is much less important for these *n*s cross sections.

It is of interest to inquire, quantitatively, as to the role of interchannel coupling in some of these satellite lines. To do this, we look at the ratios of cross sections going to a final ionic state of the same symmetry, e.g., 1$s^2$2$s$ and 1$s^2$3$s$, in a single-particle formulation with relaxation, the so-called shake-up approximation. In this approximation, for ground state photoionization to a 1$s^2$*n*s final state, the cross section

$$\sigma_{ns} \propto \langle 2s_i | ns_f \rangle \langle 2s_i | r | \varepsilon p_f \rangle \quad (5)$$

where the subscripts refer to initial (*i*) and final (*f*) state orbitals and the overlap of initial and final state 1*s* orbitals is so close to unity that it is omitted. Then looking at the ratio $\sigma_{ns}/\sigma_{2s}$ at an energy well above the thresholds, the one-electron dipole matrix element should be essentially the same for all *ns* transitions because the small difference in photoelectron energy is unimportant at high energy, as is the difference in fields experienced by the photoelectrons in the various *ns* transitions. In that case

$$\frac{\sigma_{ns}}{\sigma_{2s}} = \frac{|\langle 2s_i | ns_f \rangle|^2}{|\langle 2s_i | 2s_f \rangle|^2}, \quad (5)$$

which is independent of energy. The same considerations can be applied to the metastable state photoionization, and the ratios of the cross sections leaving the ion in the various *ns* states are also given by Eq. (5) in this approximation, and the *np* ratios are given by the same expression with all of the *ns*'s replaced by *np*'s. In any case, the ratio should be energy-independent as long as interchannel coupling among the photoionization channels is not significant. This implies that interchannel coupling is not important for the ratios $\sigma_{3s}/\sigma_{2s}$ and $\sigma_{4s}/\sigma_{2s}$ which were seen to be constant, as a function of energy, at the higher energies for both ground and excited state photoionization, and the same is true for the $\sigma_{3p}/\sigma_{2p}$ ratio in metastable state photoionization.

The results of the present shake-up calculations of the ratios using relaxed Hartree-Fock (HF) wave functions are presented in Fig. 8 and the R-matrix ratios at $E_{ph} = 8E_{range}$ are shown in Fig. 9. Comparing the ratios in the two figures, it is seen that only the ratios $\sigma_{3s}/\sigma_{2s}$ for ground-state photoionization and $\sigma_{3p}/\sigma_{2p}$ for metastable state photoionization are in good agreement with the shake-up model. This shows that the other ratios that are constant at the higher energies, and thus unaffected by interchannel coupling, are nevertheless altered by other multiparticle effects not included in the single particle shake-up model; this means that configuration interaction in the initial or final discrete state (or both) are important for these transitions. In any case, it is clear that the shake-up model is not very accurate, even for the photoionization of a four-electron system. Furthermore, agreement does not appear to get better with increasing Z. In other words, multiparticle effects persist, in general, to satellite lines in the photoionization of highly-charged ions as well.

Finally, it is of interest to note that the $\sigma_{nl}$ for the same $n$ but different $l$ show a variety of patterns as functions of energy. For ground state photoionization, $\sigma_{3s}$ and $\sigma_{4s}$ are always the largest partial cross sections in the $n=3$ and $n=4$ manifolds respectively, but the comparison between $\sigma_{np}$ and $\sigma_{nd}$ are more complicated. For $n=3$, $\sigma_{3p}<\sigma_{3d}$ is true at the higher energies for all of the ions studied. For $n=4$, $\sigma_{4p}$ is above $\sigma_{4d}$ near threshold but $\sigma_{4d}$ becomes greater than $\sigma_{4p}$ at some (Z-dependent) energy $E_1$; and at some higher (Z-dependent) energy $E_2$, they reverse again and $\sigma_{4p}>\sigma_{4d}$ at the highest energies. The calculations show that $E_1$ is about $4E_{\text{range}}$ for Be, $1.5E_{\text{range}}$ for Si$^{+10}$ and $1E_{\text{range}}$ for Fe$^{+22}$, while $E_2$ is above our energy range for Be, $4.5E_{\text{range}}$ for Si$^{+10}$ and $4E_{\text{range}}$ for Fe$^{+22}$. For the metastable state $\sigma_{nl}$, similar relations are found: $\sigma_{3p}>\sigma_{3d}>\sigma_{3s}$, $\sigma_{4p}>\sigma_{4d}>\sigma_{4s}$ in the higher energy range, and $\sigma_{4p}>\sigma_{4s}>\sigma_{4d}$ in the lower energy range. The priority of the $\sigma_{ns}$ cross sections for the ground state and $\sigma_{np}$ for the metastable state is explained qualitatively by the shake-up model in terms of the relaxation of the excited orbitals [8].

## IV. CONCLUSIONS

The partial cross sections for photoionization of ground state and metastable state Be-like ions have been calculated using the nonrelativistic *R*-matrix method; the total cross sections obtained, reported earlier [2] were in good agreement with available experiments. The photon energy range in this study is up to eight times the energy range from the ionization threshold to 4*f* threshold for each ion. The calculated partial cross sections cover the energy range high enough to represent their general behaviors well above the region dominated by autoionizing resonances, with respect to photon energy and *Z*.

To understand the relative strengths of the partial cross sections for the various satellite lines to the main lines, $\sigma_{2s}$ for the ground state and $\sigma_{2p}$ and $\sigma_{2s}$ for the metastable state, the ratios $\sigma_{nl}/\sigma_{2s}$ and $\sigma_{nl}/\sigma_{2p}$ for the ground and metastable states, respectively, were studied. These ratios are found to be smooth monotonic functions of *Z* and approach zero asymptotically at $Z\to\infty$ as predicted from general considerations. The ratios as functions of photon energy are much more complicated. A simple single-particle shake-up model failed to explain the details of most of the satellite cross sections, indicating that even in the photoionization of a four-electron system, multiparticle effects over and above relaxation are important. This leads us to infer that the behavior of satellite (photoionization plus excitation) cross sections for systems with more electrons will be at least as complicated as the present case.

## Acknowledgments

This work was supported by U.S. Department of Energy, Division of Chemical Sciences, and NSF. All calculations were performed using National Energy Research Scientific Computing Center (NERSC) computational resources.

**Figure Captions**

1. Ratios of partial cross sections $\sigma_{nl}$ to the 2s partial cross section $\sigma_{2s}$ for ground state Be against scaled photon energy.
2. As Fig. 1 but for ground state $Si^{+10}$.
3. As Fig. 1 but for ground state $Fe^{+22}$.
4. Ratios $\sigma_{nl}/\sigma_{2s}$ for all the ground state ions at the photon energy $E_{ph} = 8E_{range}$ vs. $1/Z$.
5. Ratios of partial cross sections $\sigma_{nl}$ to the 2p partial cross section $\sigma_{2p}$ for metastable state Be against scaled photon energy.
6. As Fig.5 but for metastable state $Si^{+10}$.
7. As Fig. 5 but for metastable state $Fe^{+22}$.
8. Relaxed Hartree-Fock (HF) values of $|\langle P_{2s}^{(i)}|P_{ns}^{(f)}\rangle / \langle P_{2s}^{(i)}|P_{2s}^{(f)}\rangle|$ for ground state photoionization and $|\langle P_{2p}^{(i)}|P_{np}^{(f)}\rangle / \langle P_{2p}^{(i)}|P_{2p}^{(f)}\rangle|$ and $|\langle P_{2s}^{(i)}|P_{ns}^{(f)}\rangle / \langle P_{2s}^{(i)}|P_{2s}^{(f)}\rangle|$ for metastable state photoionization vs. Z, where (*i*) and (*f*) indicate the radial functions in the initial and final state HF wave functions respectively. In the legend "Gr" denotes ground state, "Ex" denotes metastable state, and "$nl/2l$" means the value of $|\langle P_{2l}^{(i)}|P_{nl}^{(f)}\rangle / \langle P_{2l}^{(i)}|P_{2l}^{(f)}\rangle|$.
9. As Fig. 8 except that the ratios are from the R-Matrix calculation. The legend is the same as Fig. but "$nl/2l$" stands for $\sigma_{nl}/\sigma_{2l}$.

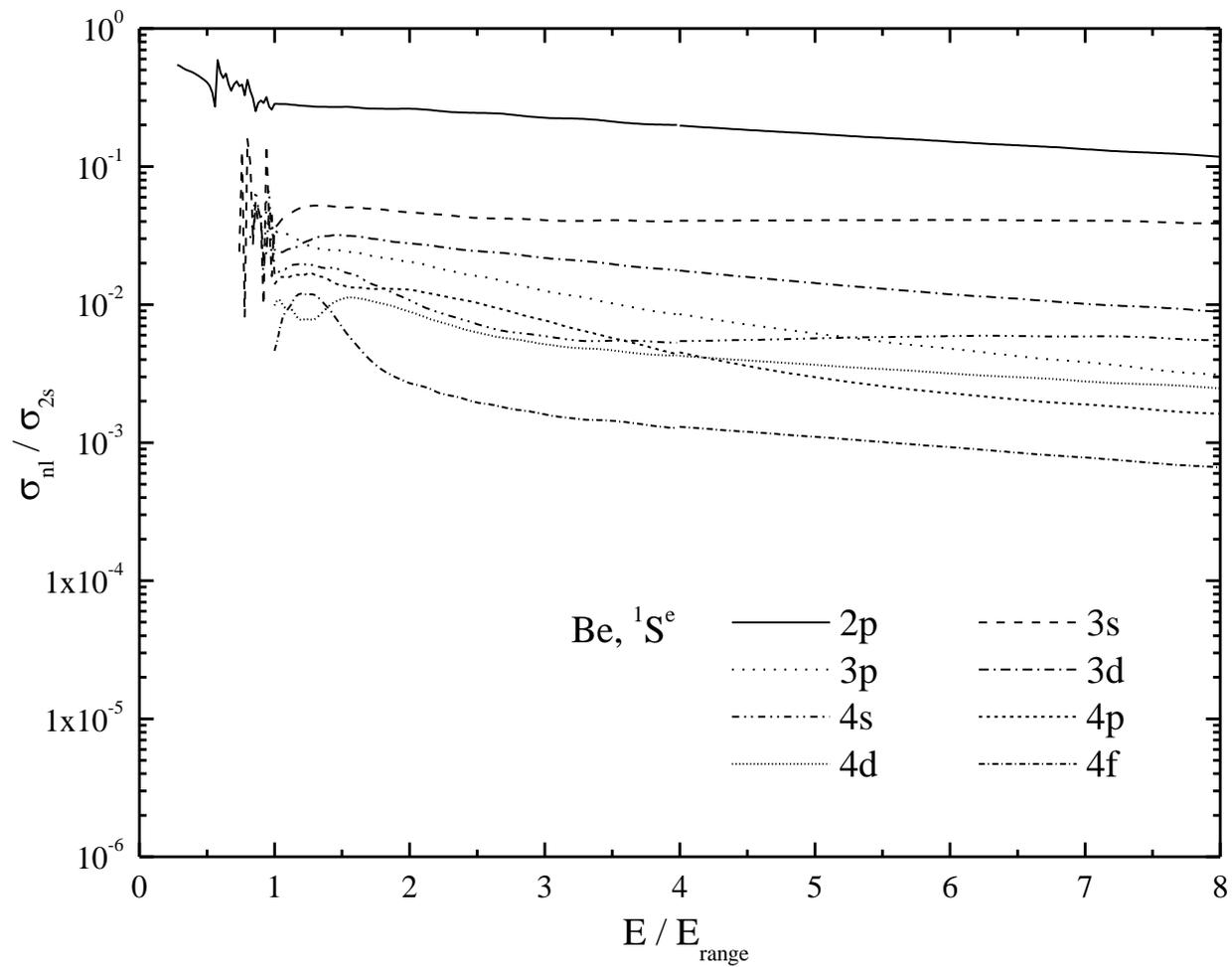

Fig. 1

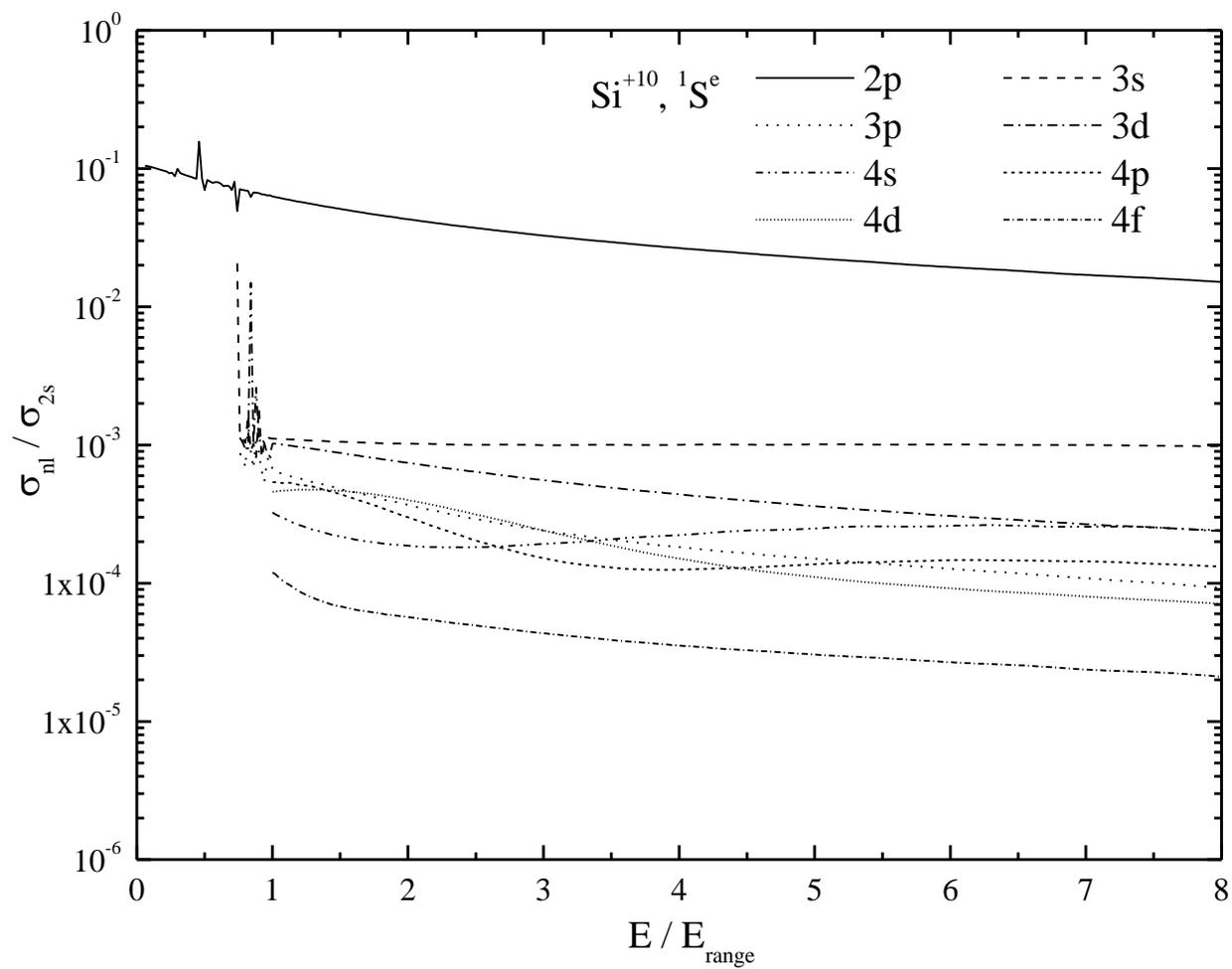

Fig. 2

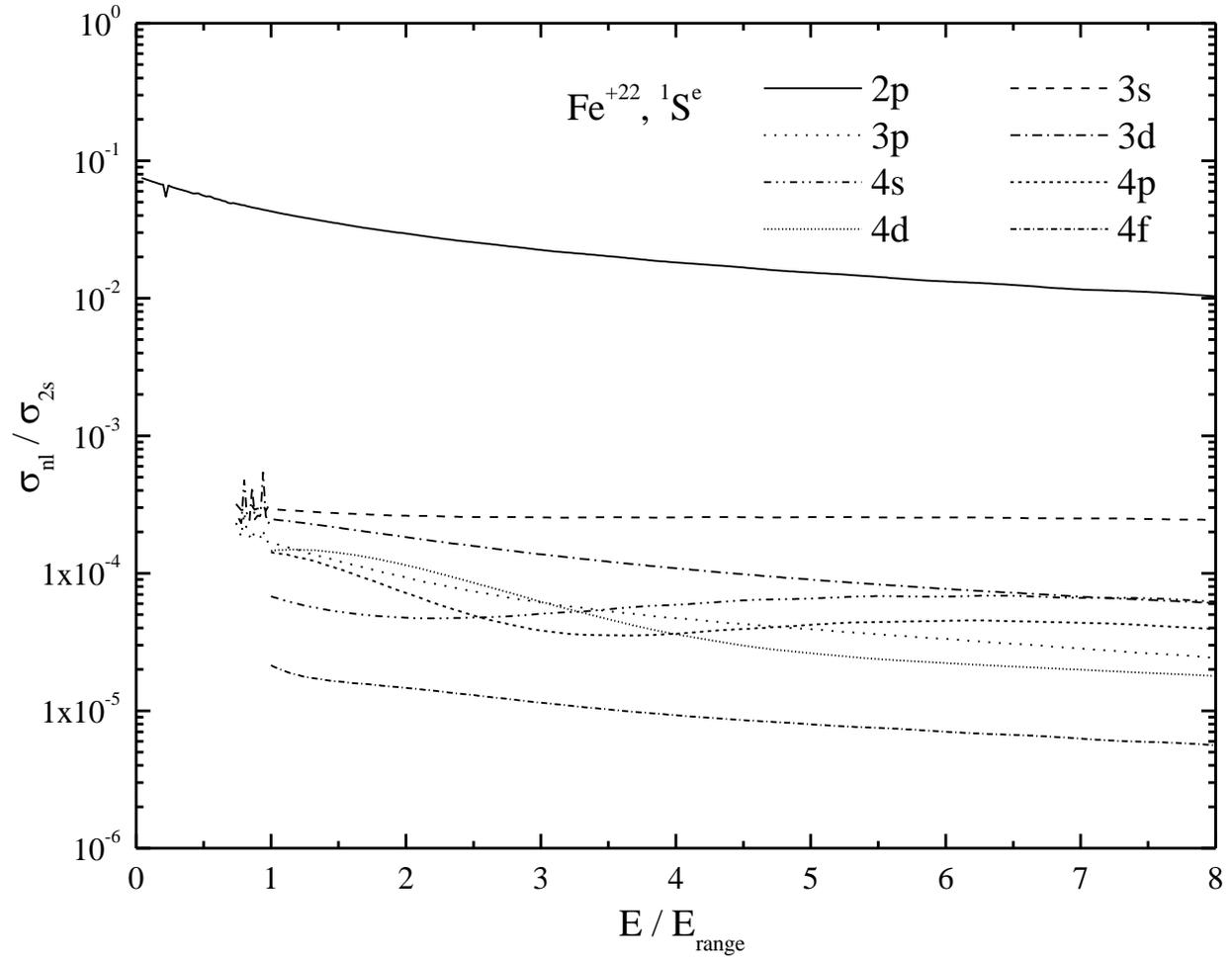

Fig. 3

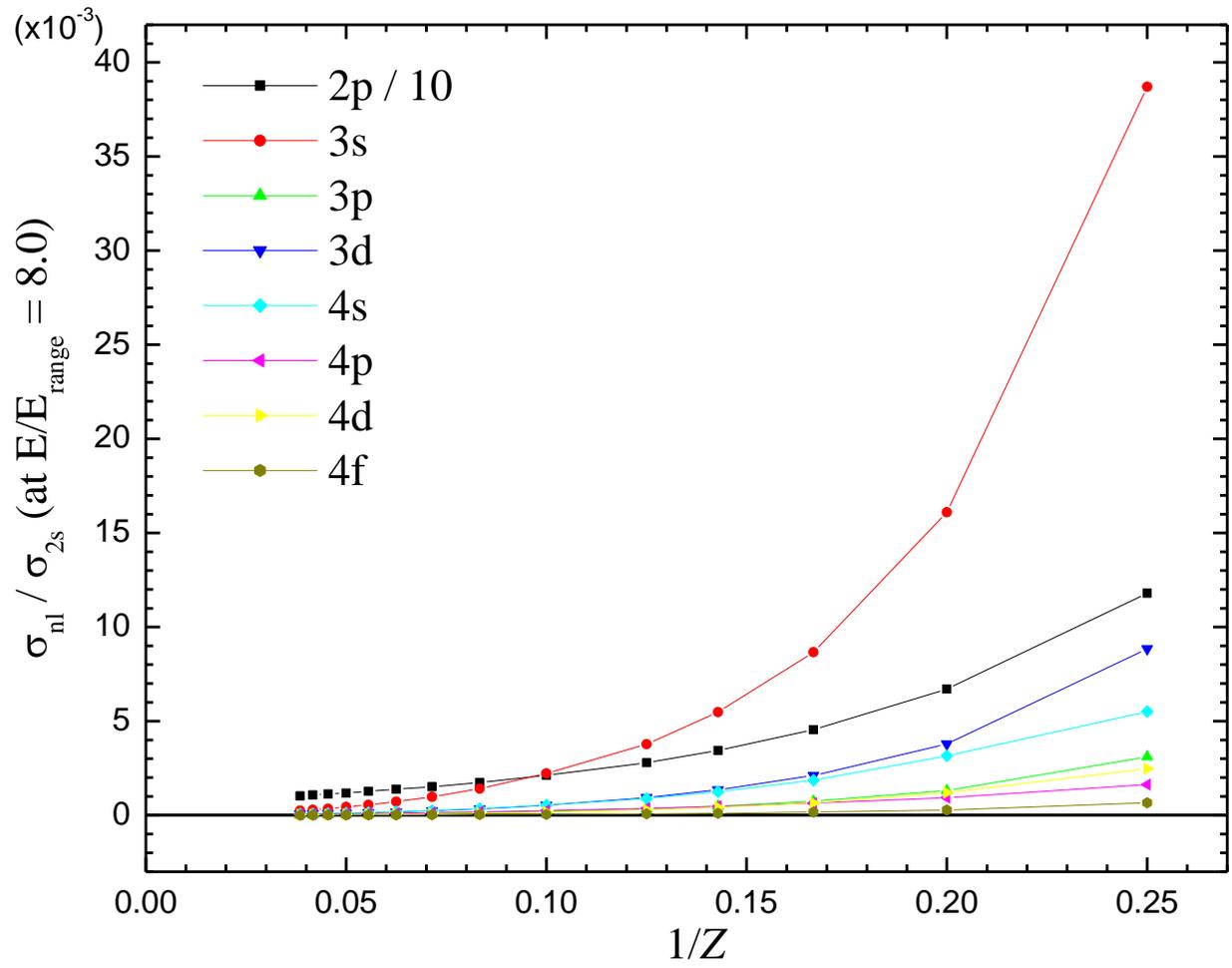

Fig. 4

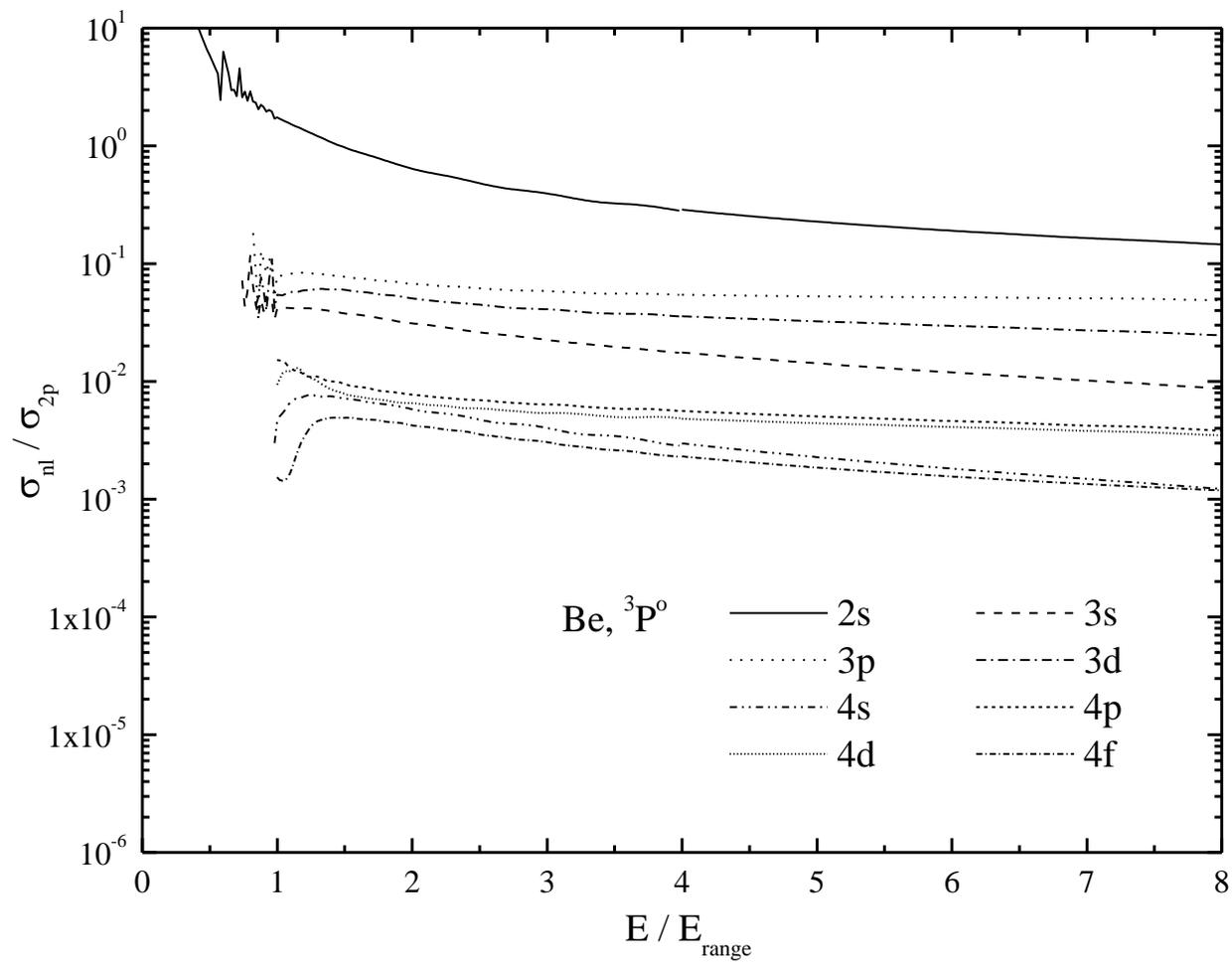

Fig. 5

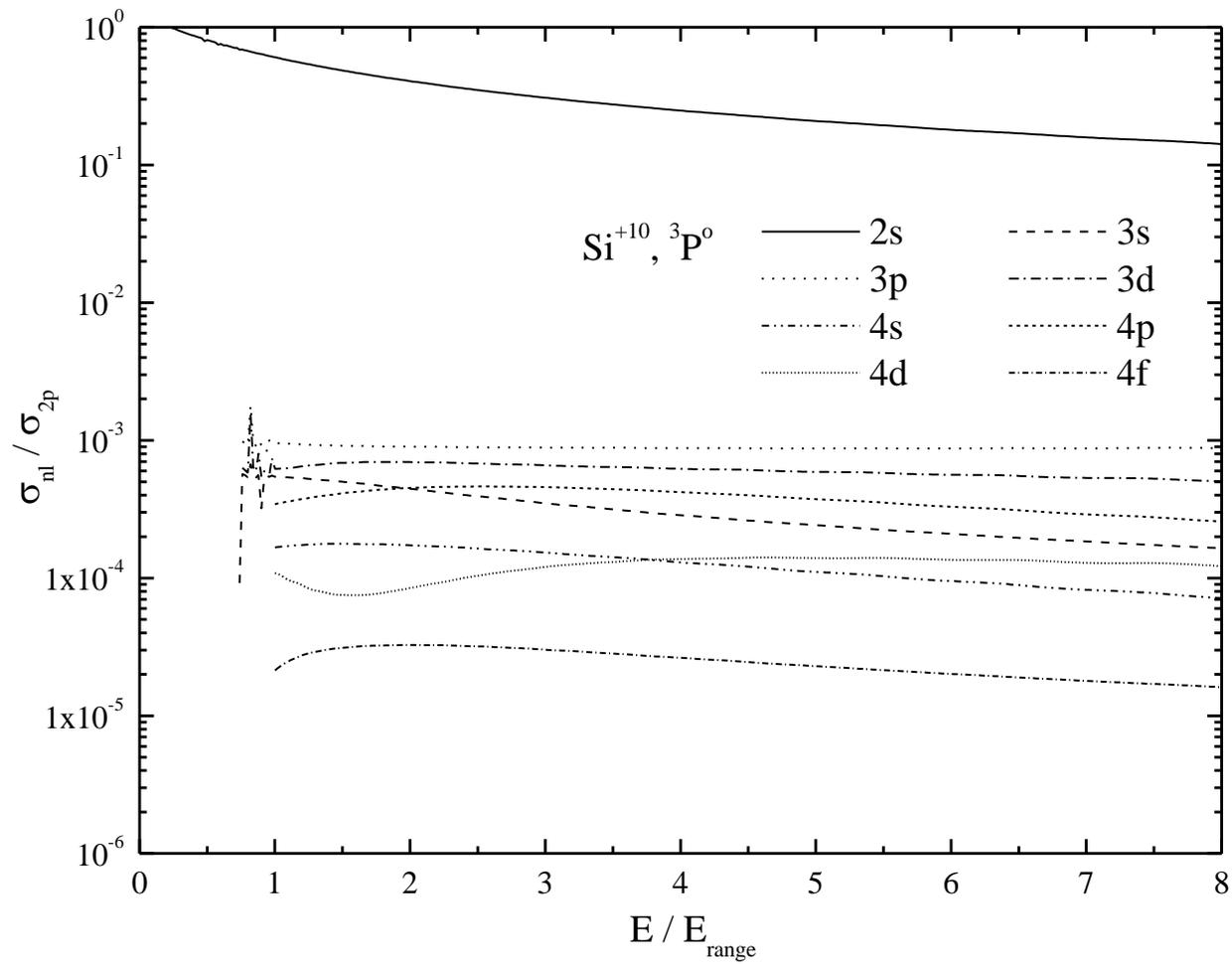

Fig. 6

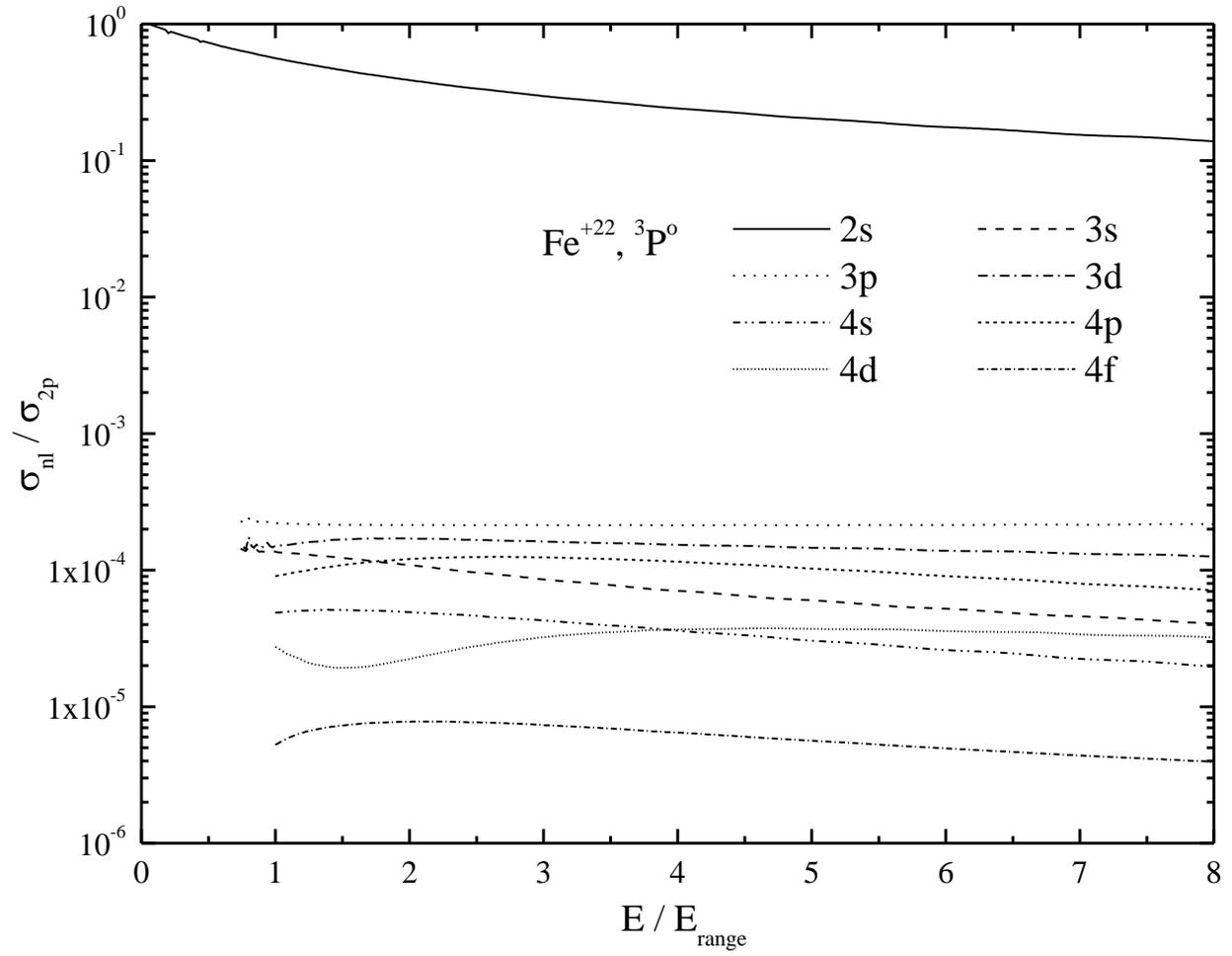

Fig. 7

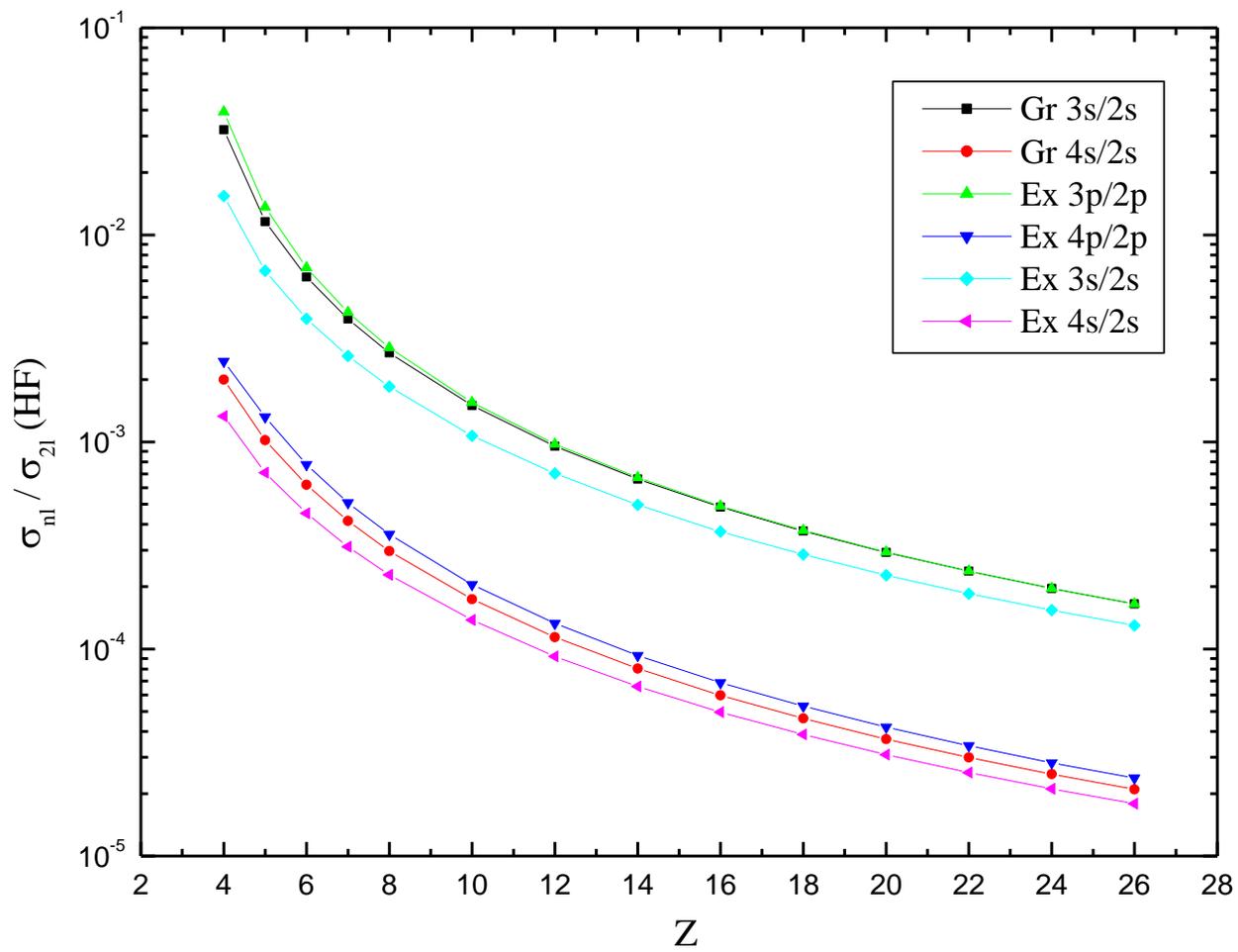

Fig. 8

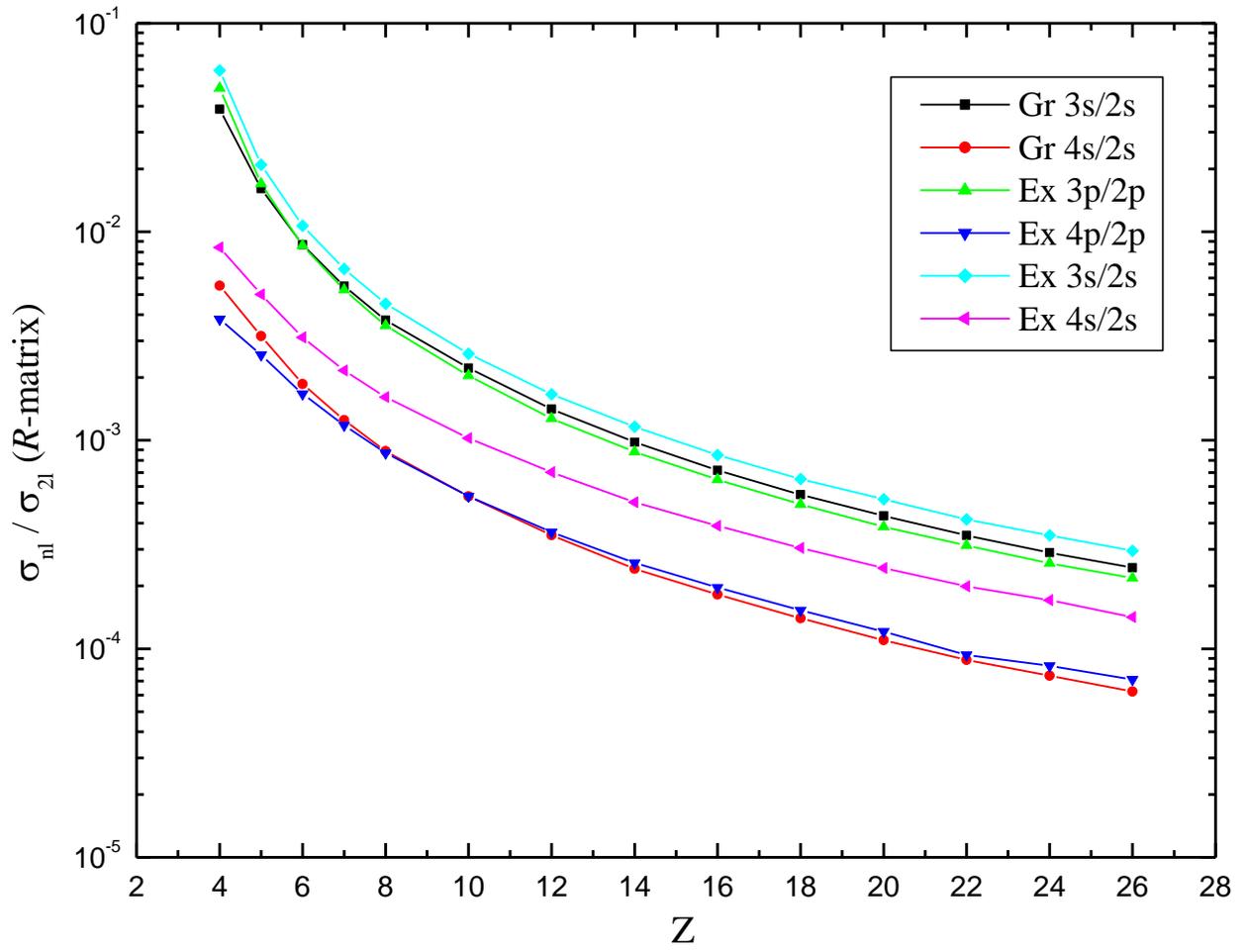

Fig. 9